# A Universe with both Acceleration and Deceleration

Nalin de Silva

Department of Mathematics, University of Kelaniya, Kelaniya, Sri Lanka

## 1. INTRODUCTION

Since Perlmutter and others (1997) & (1998) showed that the universe expands with an acceleration, many models have been proposed to explain this phenomenon. Dark energy and Einstein's cosmological constant have been prominent in the models that have been suggested. With the announcement of the observations on deceleration at an earlier epoch of the universe by Schaefer (2006) it has been suggested that the cosmological constant has to be ruled out, as a constant cannot give both acceleration and deceleration. However it is possible to obtain models with both acceleration and deceleration with a variable, which we have been studying recently. It has to be mentioned that a variable was first introduced by us in order to explain the acceleration of the universe, before the deceleration of the universe was observed by Schaefer.

## 2. MODIFIED FIELD EQUATIONS

A variable has been studied by many authors, reference to whom could be found in Harko and Mak (1999) who considered particle creation in cosmological models with varying gravitational and cosmological "constants". We vary only the cosmological "constant" with suitable modifications in the Einstein field equations. We write the field equations in the form $G^{\mu\nu} = \kappa T^{\mu\nu} + \Lambda g^{\mu\nu}$, where $\kappa$ is a constant and $\Lambda$ is a variable, and taking into consideration of the fact that $G^{\mu\nu}$ is divergenceless, impose the condition that the entire right hand side of the equation is divergenceless, instead of the condition that $T^{\mu\nu}$ is divergenceless. Then it is not the energy momentum of matter and radiation that is conserved, but the energy momentum of matter, radiation and the energy of the field.

Using the modified field equations, it can be shown that the density $\rho$, the pressure $p$ of the universe, $\Lambda(t)$, and $R(t)$ in the Robertson Walker metric satisfy the equations

$$\kappa p + \frac{kc^2}{R^2} + \frac{2\ddot{R}}{R} + \frac{\dot{R}^2}{R^2} = \Lambda$$

$$\kappa \rho + \frac{\Lambda}{c^2} = \frac{3k}{R^2} + \frac{3\dot{R}^2}{R^2 c^2}.$$

where . denotes differentiation with respect to cosmic time t.

These two equations lead to $3\left(\rho + \frac{p}{c^2}\right)\frac{\dot{R}}{R} + \dot{\rho} + \frac{\dot{\Lambda}}{\kappa c^2} = 0$.

The above two equations can be solved for $R$, $p$, $\rho$ and $\Lambda$, for a given value of $k$. As there are four variables and only two equations, no unique solution is possible, and one could study different families of solutions, and finally select a particular solution that satisfies the data from observations.

3. FAMILY OF SOLUTIONS

Hemantha and de Silva (2003) have considered solutions of the form $R = a + b_1 \sin\ t$, and Rahubedde and de Silva (2005) have studied solutions of the form $R = a + b_1 \cos\ t + b_2 \cos\ 2\ t$. However, what is discussed in this paper is based on the family of solutions in the form $R = a + b_1 \cos\ t + b_3 \cos 3\ t$, described by Hemantha and de Silva (2004).

Taking $k = 1$, $p = 0$, $R(t) = a + b_1 \cos \omega t + b_3 \cos 3\omega t$, one finds that

$$\Lambda(t) = 0.75\left[\frac{2c^2}{(a + b_1 \cos\omega t + b_3 \cos 3\omega t)^2} - 2\omega^2\left(\frac{b_1 \cos\omega t + 9b_3 \cos 3\omega t}{a + b_1 \cos\omega t + b_3 \cos 3\omega t}\right) + 1.5\omega^2\left(\frac{b_1 \sin\omega t + 3b_3 \sin 3\omega t}{a + b_1 \cos\omega t + b_3 \cos 3\omega t}\right)^2\right]$$

where $a, b_1$ and $b_3$ are constants.
R could be schematically represented by

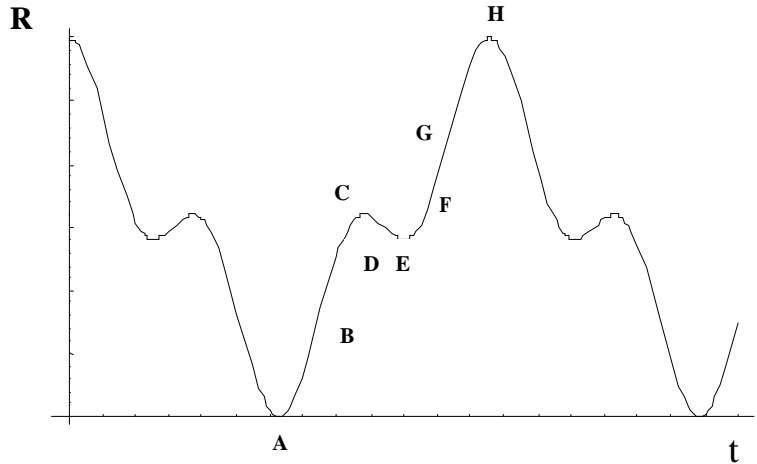

The figure shows that the universe expands and contracts in cycles, and in the present cycle, the universe expands during phase AB with an acceleration then in BC with deceleration, and contracts with a deceleration in CD, before acceleration sets in at D. The universe expands with an acceleration from E to G, and with a deceleration from G to H. The present epoch is represented by F. As no blue shifts have been observed so far the value of $R$ at F should be greater than or equal to that at C, the maximum value that $R$ has attained in the past in the present cycle of the universe. The acceleration of the universe observed by Perlmutter et.al. could correspond to the phase DEF, and the deceleration claimed by Schaefer could be represented by the phase BCD.

The universe accelerates when the contribution from the field is more than that due to matter and radiation. Initially at A, the field dominates, but however, as the universe expands, some of the energy from the field is converted to matter and radiation. As a result at B, field balances with the matter and radiation field, and thereafter deceleration sets in. However, again at D the field begins to dominate when the universe is given a second accelerating phase. This phase continues up to G, when as a result of conversion of energy from the field to matter and radiation, a second decelerating phase sets in. It has to be pointed out that the total energy of the field, and of matter and radiation is conserved. Cyclic models have been studied by others including Steinhardt and Turok (2002) & (2004) in a different context.

It is easily seen that $\dot{R}$ becomes zero when $t$ is zero or $\cos\;t = ((3b_3 - b_1)/12\,b_3)$. In order to satisfy the latter condition we demand that $b_1 < 3\,b_3$. We also find that $\ddot{R} = 0$ when $\cos 3\;t = ((27b_3 - b_1)/36\,b_3)$. Thus $\dot{R} = 0$ when $t = $ , $+$ , $2 -$ , $2$ , where $\cos = ((3b_3 - b_1)/12\,b_3)$, in half of a cycle. The points A, C, E and H in the figure correspond to , $+$ , $2 -$ and $2$ respectively. $\ddot{R} = 0$ when $t = $ $+$ , $3 /2, 2 -$ , with $\cos = ((27\,b_3 - b_1)/36\,b_3)$, corresponding to the points B, D and G respectively in the same half cycle. It should be noted that $<$ .

We take the value of $R$ at  to be zero making $a = b_1 + b_3$. The present acceleration of the universe starts at D corresponding to $t = 3\pi/2$. Let the ratio of the value of $R$ at the present epoch to that at $3\pi/2$ be equal to $q$, and the ratio of the value of $R$ at the present epoch to that at B, where the deceleration in the present cycle begins be equal to $p$. Thus $p$ is the redshift at which the deceleration begins after the initial acceleration, and $q$ is the redshift at which the second accelerating phase begins. These relationships lead us to the equation $x^3 + (\alpha^2 - 27)x^2 + 2\alpha^2 x + \alpha^2 = 0$, where $\alpha = 27(p-q)/4p$ and $x = b_1/b_3$. This equation has two real roots if $0 < \alpha < r$, where $r$ is about 3.93. Taking $\alpha = 3.93$ we find $x=6$, giving $b_1 = 6b_3$. However, we have found that $b_1 < 3b_3$ or $x < 3$, and therefore $b_1=6b_3$ is inadmissible. We take $x = 2.99$ or $b_1=2.99b_3$, which gives $\alpha = 3.67$. If we take $q = 1.6$, the value generally quoted for the redshift at which the second acceleration began we obtain $p = 3.5$ for the redshift at which deceleration began. This value may not necessarily be the value given by Schaefer making use of the gamma ray bursts. However the model at least gives a value for the onset of deceleration, and it may be possible to improve the model varying some of the parameters.

With $b_1=2.99 b_3$, $a$ becomes equal to $3.99 b_3$, and we can express $R$ in terms of $b_3$ and . Also we find that for the above values $\cos\beta = 0.03$ and $\cos\gamma = 0.82$. Using these values $R$ can be calculated at B, C, D, E, G and H, in terms of $b_3$. For example $R$ at C, D and E are respectively given by $3.99019 b_3$, $3.99 b_3$ and $3.9898 b_3$. It should be pointed out that the value of $R$ at G is $6.1608 b_3$ which is very much larger than the values at C, D and E. As $R$ changes slowly from C to E, we take its value at F, the present epoch, to be $3.99026 b_3$, slightly greater than its value at C. This is an arbitrary value chosen to illustrate the main argument of an accelerating universe with no blue shifts, in the present epoch. This value corresponds to $\omega t = 3\pi/2 + 0.0192$, and assuming that the age of the universe is 13.7 billion years in the present cycle, we obtain $\omega = 1.1051* 10^{-17}$ radians per second.

Finally it can be shown that, the density is always positive, after eliminating  from the first two equations, for pressure free models with $k = 1$. It can also be shown that when $p = 0$, the rate of change of energy of matter and radiation per unit energy of matter and radiation $\dfrac{\frac{d}{dt}(\rho R^3)}{\rho R^3} = \dfrac{\dot\rho}{\rho} + \dfrac{\dot R}{R} \cong -\dfrac{\dot R}{R}$, for the values for the present epoch, which is very small, as it is of order . Thus there is no violation of the law of conservation of energy of matter and radiation, at the present epoch, within the experimental limits.

4. CONCLUSIONS

It is seen that the family of solutions discussed above, in terms of the parameter $b_3$, gives solutions of the modified field equations, that represent the universe with a decelerating phase after the initial accelerating phase, and a second accelerating phase before a second decelerating phase sets in.